\documentclass[conference]{IEEEtran}
\IEEEoverridecommandlockouts
\usepackage{cite}
\usepackage{amsmath,amssymb,amsfonts}
\usepackage{algorithmic}
\usepackage{graphicx}
\usepackage{textcomp}
\usepackage{array}
\usepackage{booktabs}
\usepackage{xcolor}
\def\BibTeX{{\rm B\kern-.05em{\sc i\kern-.025em b}\kern-.08em
    T\kern-.1667em\lower.7ex\hbox{E}\kern-.125emX}}
    
\begin{document}


\title{Energy-Efficient Memristor Accelerator for Depression Detection with a CNN-LSTM Hybrid Model}

\author{
\IEEEauthorblockN{
Jiaqi Liu\textsuperscript{1,$\dagger$},
Yun Yu\textsuperscript{1,$\dagger$},
Fenglin Xi\textsuperscript{1,$\dagger$},
Shenjian Zhang\textsuperscript{1},
Shenghao Liao\textsuperscript{1},
Wenze Zheng\textsuperscript{1},\\
Mingxuan Hu\textsuperscript{1},
Yangbin Chen\textsuperscript{1,*},
Xinhong Chen\textsuperscript{1,*},
Chun Zhao\textsuperscript{1,2,*}
}
\IEEEauthorblockA{
\textsuperscript{1}
\textit{School of Advanced Technology,}
\textit{Xi'an Jiaotong-Liverpool University,}
Suzhou, China
}
\IEEEauthorblockA{
\textsuperscript{2}
\textit{HC Cloudminds Microelectronics Technology Co., Limited}
}
\IEEEauthorblockA{
Email: Jiaqi.Liu03@xjtlu.edu.cn
}
\IEEEauthorblockA{
\textsuperscript{$\dagger$}These authors contributed equally to this work
and share first authorship.
}
\IEEEauthorblockA{
\textsuperscript{*}Corresponding authors:
Yangbin Chen, Xinhong Chen, and Chun Zhao.
}
}
\vspace{-20pt}

\maketitle
\begin{abstract}
Early screening of depression requires efficient analysis of audio and
speech data, while conventional CPU and GPU platforms incur high latency
and energy consumption for CNN-LSTM inference. To address this challenge,
we propose a unified resistive RAM (ReRAM) processing-in-memory accelerator
for a CNN-LSTM attention workload. The architecture integrates
convolutional, recurrent, attention, and multi-layer perceptron (MLP) stages within a crossbar-based
framework, reducing intermediate data movement between spatial and temporal
branches. CNN operations are mapped through im2col-based vectorization,
while LSTM gate computations are reformulated using concatenated input-state
vectors and mapped onto ReRAM crossbars. We further provide module-level
workload breakdown, parameter sensitivity analysis, and matched-workload
context against representative ReRAM/PIM accelerators. Experimental results
show 32.1 times and 9.3 times latency reduction over CPU and GPU baselines,
respectively, with about 160 times and 140 times energy reduction.
Simplified weight-perturbation tests characterize ASLF sensitivity in an
untrained pipeline with synthetic spectrogram-shaped inputs. Task-level
prediction accuracy and robustness remain to be evaluated.

\end{abstract}

\begin{IEEEkeywords}
ReRAM accelerator, processing-in-memory, CNN-LSTM, depression detection, device nonideality.
\end{IEEEkeywords}

\section{Introduction}

Depression is a major global health issue, affecting a large population
worldwide and often remaining undiagnosed until severe stages. Early
screening is important because timely intervention can improve treatment
outcomes and reduce long-term societal and economic burdens. Recently,
deep neural network (DNN)-based methods have shown strong potential for
automatic depression detection, providing objective and scalable
alternatives to traditional clinical assessment~\cite{niu2020multimodal}.

Hybrid models combining convolutional neural networks (CNNs) and long
short-term memory networks (LSTMs) are particularly suitable for
speech-based depression detection. CNNs extract local time--frequency
patterns from spectrograms, while LSTMs capture temporal dependencies
across speech frames. However, such CNN--LSTM pipelines are
computationally expensive: convolutional layers require intensive
matrix operations, and LSTM gates must be evaluated sequentially across
time steps. Executing these workloads on CPUs and GPUs therefore incurs
high latency and energy consumption, limiting real-time deployment in
resource-constrained scenarios~\cite{sze2017efficient}.

Processing-in-memory (PIM) architectures based on resistive RAM (ReRAM)
provide an attractive solution by executing matrix--vector
multiplications close to where weights are stored. Prior ReRAM
accelerators have demonstrated high efficiency for neural-network
inference, including CNN-oriented, pipelined, and recurrent workloads
~\cite{rram_survey,prime2016,isaac,pipelayer,eralstm}. Nevertheless,
many existing designs mainly target individual workload types. A
speech-based CNN--LSTM attention model requires not only convolutional
and recurrent computation, but also feature transfer, attention fusion,
and final MLP processing within a unified dataflow.

To address this need, we propose a unified ReRAM-based accelerator for a
CNN--LSTM attention pipeline in depression detection. The architecture uses heterogeneous CNN and LSTM tiles connected through
a hierarchical network-on-chip (NoC), with fixed-weight fully connected (FC) operations mapped onto ReRAM crossbars and input-dependent attention handled by on-chip digital vector units and special function units (SFUs) in the evaluated dataflow. The design supports spatial feature extraction,
temporal modeling, audio segment-level feature (ASLF) representation extraction, and final prediction
within one ReRAM-based framework.

The main contributions are summarized as follows:
\begin{itemize}
    \item We propose a unified ReRAM-based accelerator for hybrid
    CNN--LSTM inference, integrating CNN tiles, LSTM tiles, NoC-based
    feature transfer, attention fusion, and final MLP processing.
    \item We reformulate CNN, LSTM, and FC operations for ReRAM
    crossbar execution, while input-dependent attention is handled by
    on-chip digital vector units and nonlinear operations by SFUs.
    \item We provide module-level workload breakdown, parameter
    sensitivity analysis, and matched-workload context with representative
    ReRAM/PIM accelerators, including ISAAC, ERA-LSTM, and PUMA.
    \item We characterize ASLF-level sensitivity to simplified
    weight variation and drift in an untrained pipeline with synthetic
    inputs, without inferring task-level prediction robustness.
\end{itemize}

\section{PRELIMINARY}

\subsection{CNN--LSTM Attention Workload}

\begin{figure}[ht]
    \centering
    \includegraphics[width=0.5\linewidth]{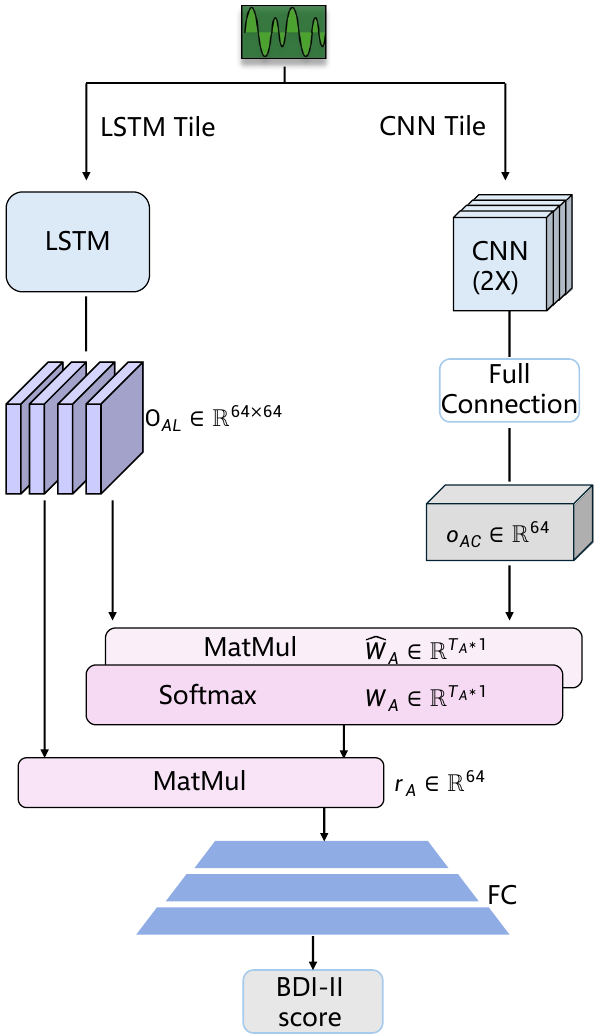}
    \caption{A CNN-LSTM attention pipeline for ASLF representation extraction and BDI-II score prediction.}
    \label{fig:depression}
\end{figure}

The target workload is adapted from the audio spatio-temporal attention
(STA) network of Niu et al.~\cite{niu2020multimodal}. It combines the
spectrogram-based CNN/LSTM feature-extraction and attention structure with
the regression head defined below. The input spectrogram has 129 frequency
bins and 64 temporal frames. As shown in Fig.~\ref{fig:depression}, the
model extracts spatial and temporal depression-related cues through two
complementary processing paths.

\textbf{CNN path.} 
The CNN branch extracts local time--frequency patterns from the input 
spectrogram. Two convolutional layers with $5 \times 11$ kernels and 
stride $(3,3)$ are applied, followed by a fully connected layer with 
64 neurons. The CNN branch produces a 64-dimensional spatial feature 
vector:
\begin{equation}
    O_{AC} = \mathrm{ReLU}(W_{FC} O_3 + b_{FC}), 
    \quad O_{AC} \in \mathbb{R}^{64}.
\end{equation}
Here, $O_3$ denotes the flattened CNN feature map supplied to the
fully connected layer.

\textbf{LSTM path.} 
The spectrogram columns are treated as a sequence of 64 time steps, 
where each time step is represented by a 129-dimensional input vector. 
The LSTM branch captures temporal dependencies across the speech segment 
and generates a sequence of hidden states:
\begin{equation}
    O_{AL} = [h_1, h_2, \ldots, h_T]^T, 
    \quad O_{AL} \in \mathbb{R}^{T \times 64}, 
    \quad T = 64.
\end{equation}

\textbf{Attention fusion.} 
The CNN spatial feature is used to attend to the temporal features 
generated by the LSTM branch. The attention score is computed by a 
matrix--vector multiplication between the LSTM feature matrix and the 
CNN feature vector:
\begin{equation}
    s_A = O_{AL} O_{AC}, 
    \quad s_A \in \mathbb{R}^{T}.
\end{equation}
The attention weights are obtained by a softmax operation:
\begin{equation}
    w_A = \mathrm{softmax}(s_A), 
    \quad w_A \in \mathbb{R}^{T}.
\end{equation}
The fused representation is then computed by weighted aggregation over 
the temporal features:
\begin{equation}
    r_A = O_{AL}^{T} w_A, 
    \quad r_A \in \mathbb{R}^{64}.
\end{equation}

\textbf{ASLF representation and final prediction.}
The fused representation $r_A$ is processed by a three-layer fully
connected network. The first two FC layers generate a 64-dimensional
audio segment-level feature (ASLF), while the last FC layer produces the
final scalar prediction:
\begin{equation}
    R_1 = \mathrm{ReLU}(W_{FC1} r_A + b_{FC1}),
    \quad R_1 \in \mathbb{R}^{64},
\end{equation}
\begin{equation}
    ASLF = \mathrm{ReLU}(W_{FC2} R_1 + b_{FC2}),
    \quad ASLF \in \mathbb{R}^{64},
\end{equation}
\begin{equation}
    \hat{y} = W_{FC3} ASLF + b_{FC3},
    \quad \hat{y} \in \mathbb{R}^{1}.
\end{equation}
In this work, ASLF denotes the 64-dimensional output of FC2, followed by
a scalar linear FC3 regression head. This placement and prediction head
adapt the original audio STA network, in which ASLF is extracted after a
third 64-neuron FC layer~\cite{niu2020multimodal}. The accelerator
evaluation covers the CNN, LSTM, attention, and final MLP stages, while
the non-ideality analysis is reported at the ASLF representation level
rather than as final task-level BDI-II prediction accuracy under device
perturbations.

Mapping this algorithm to the proposed accelerator is efficient because 
the dominant operations, including convolution, LSTM gate computation, 
attention score generation, weighted aggregation, and fully connected 
layers, can all be formulated as matrix--vector multiplications (MVMs). 
The fixed-weight CNN, LSTM, and FC operations are mapped to ReRAM
crossbar arrays~\cite{rram_survey,prime2016,isaac}. Input-dependent
attention MVMs are handled by on-chip digital vector units in the
evaluated dataflow. Nonlinear operations such as ReLU, sigmoid, tanh,
and softmax are handled by SFUs.

\subsection{MVM Reformulation for ReRAM Mapping}

\begin{figure}[ht]
    \centering
    \includegraphics[width=\linewidth]{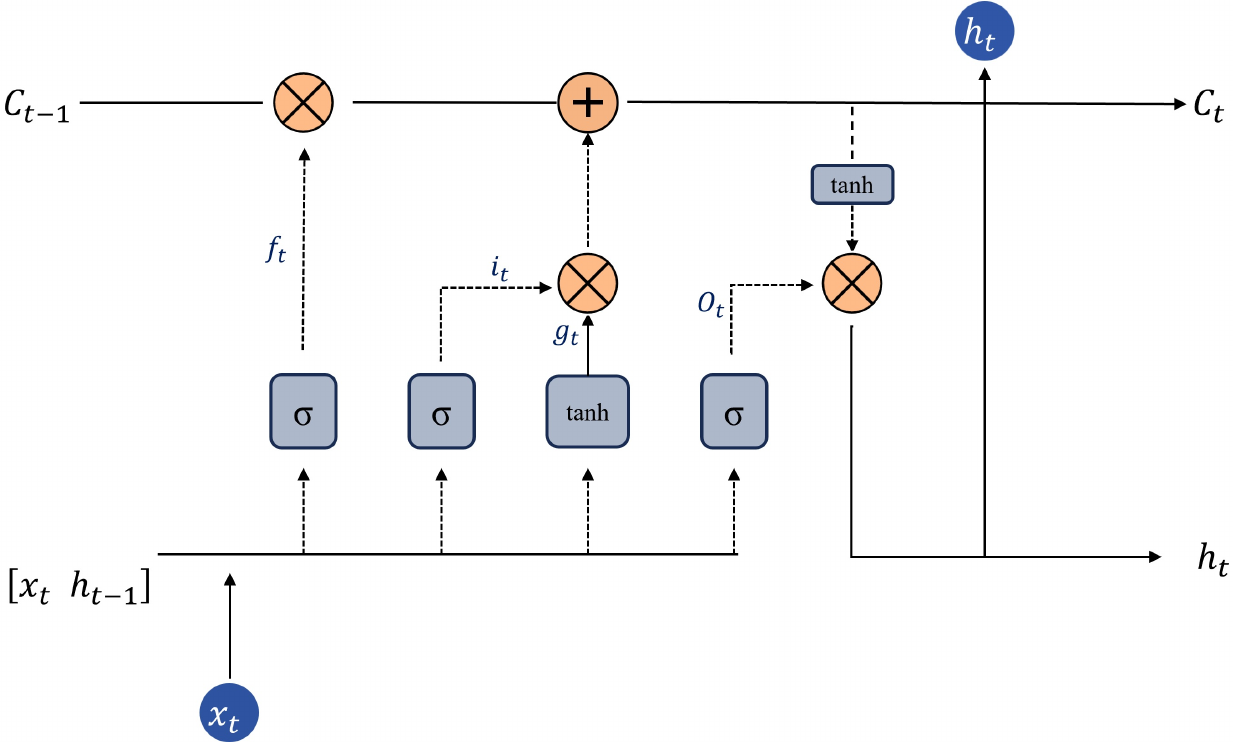}
    \caption{Structure of an LSTM cell with gated operations.}
    \label{fig:lstm_gate}
\end{figure}

ReRAM crossbar arrays are well suited for MVM operations~\cite{rram_survey,prime2016,isaac}. 
When input activations are applied as voltages through DACs and weights 
are stored as cell conductances, the output current on each bitline 
naturally performs analog multiply--accumulate operations:
\begin{equation}
    I_j = \sum_i V_i G_{ij},
\end{equation}
where $V_i$ is the input voltage and $G_{ij}$ is the programmed 
conductance of the ReRAM cell. The output currents are then digitized 
by ADCs and processed by peripheral digital units.

For convolutional layers, the input feature maps are transformed into 
vectors using the image-to-column (im2col) method~\cite{isaac,pipelayer}. 
The convolution kernels are mapped as crossbar conductances, and each 
input vector is applied through DACs. The resulting bitline currents are 
digitized by ADCs and accumulated by peripheral shift-and-add units to 
generate output feature maps.

Fig.~\ref{fig:lstm_gate} illustrates the gated computation structure of 
an LSTM cell. For LSTM layers, each gate is originally computed as:
\begin{equation}
    g_t = \phi(Ux_t + Wh_{t-1}),
\end{equation}
where $x_t$ is the input vector at time step $t$, $h_{t-1}$ is the 
previous hidden state, and $\phi(\cdot)$ denotes the corresponding 
nonlinear activation. Following the common ReRAM-based LSTM mapping 
strategy~\cite{eralstm}, the input vector and the previous hidden state 
are concatenated:
\begin{equation}
    v_t = [x_t; h_{t-1}],
\end{equation}
and the input and recurrent weight matrices are concatenated as:
\begin{equation}
    M = [U \; W].
\end{equation}
The gate computation can then be reformulated as a single MVM:
\begin{equation}
    Mv_t = Ux_t + Wh_{t-1}.
\end{equation}
This reformulation allows LSTM gate operations to be mapped efficiently 
onto ReRAM crossbars. The hidden and cell states are stored in local 
buffers and reused across time steps, while sigmoid and tanh operations 
are implemented using SFUs.

The attention fusion stage is also MVM-dominated. The attention score
$s_A = O_{AL}O_{AC}$ and weighted aggregation $r_A = O_{AL}^{T}w_A$
are computed by on-chip digital vector units in the evaluated dataflow.
Because $O_{AL}$ contains input-dependent activations, this avoids
programming a new attention matrix into ReRAM for each inference.
The softmax function is handled by SFUs. The fixed weights of the
fully connected layers for ASLF representation extraction and final
prediction are mapped onto ReRAM crossbars. Module-specific dataflow
control coordinates these analog and digital operations.

\section{Architecture Design}

In this section, we present the proposed ReRAM-based accelerator
architecture for the CNN--LSTM attention workload. We first describe the
overall architecture and hybrid dataflow, then introduce the
micro-architectures of CNN and LSTM tiles, and finally explain the
mapping of attention and fully connected (FC) operations.

\begin{figure*}[t]
    \centering
    \includegraphics[width=\textwidth]{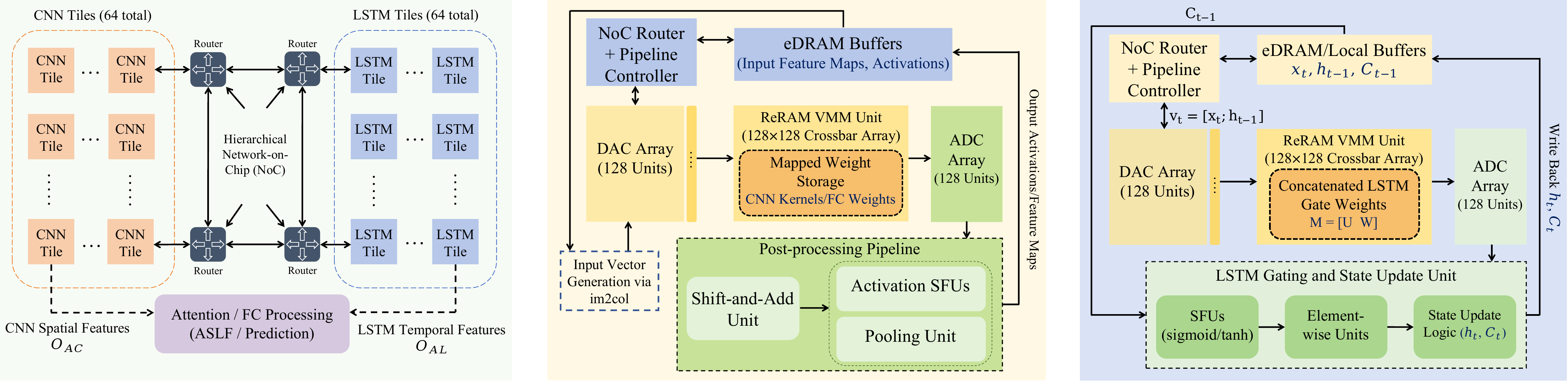}
    \caption{Proposed ReRAM-based accelerator architecture for the CNN--LSTM attention workload.
    (a) System-level organization and NoC-based hybrid dataflow.
    (b) CNN tile micro-architecture.
    (c) LSTM tile micro-architecture.}
    \label{fig:architecture}
\end{figure*}

\subsection{Overall Architecture and Hybrid Dataflow}

Fig.~\ref{fig:architecture} illustrates the overall architecture of
the proposed depression detection accelerator. The system is composed of
heterogeneous CNN tiles and LSTM tiles connected through a hierarchical
network-on-chip (NoC). CNN tiles are responsible for convolutional and
fully connected computations in the spatial feature extraction path,
while LSTM tiles accelerate recurrent computations for temporal
modeling. The extracted CNN spatial feature and LSTM temporal features
are then transferred through the NoC to the attention/FC processing
stage for ASLF representation extraction and final prediction.

Each tile is built around ReRAM-based vector--matrix multiplication
(VMM) units, together with DACs, ADCs, local eDRAM buffers, routing
logic, pipeline controllers, and special function units (SFUs). This
organization follows the general ReRAM processing-in-memory principle,
where the dominant multiply--accumulate operations are executed close to
where the weights are stored, thereby reducing memory access and data
movement overhead~\cite{rram_survey,prime2016,isaac}.

The proposed architecture is designed for the hybrid CNN--LSTM workload
instead of treating CNN and LSTM as two isolated accelerator pipelines.
The input spectrogram is processed in parallel by the CNN and LSTM
branches. The CNN branch generates a compact spatial feature vector
$O_{AC}$, while the LSTM branch produces a temporal feature matrix
$O_{AL}$. These intermediate features are transferred through the NoC to
the attention/FC processing stage, enabling on-chip fusion between
spatial and temporal representations. By coordinating feature extraction
and fusion within a unified architecture, the proposed design reduces
intermediate data movement and supports the end-to-end CNN--LSTM
inference flow more efficiently.

\subsection{CNN and LSTM Tile Micro-architectures}

The proposed accelerator employs two complementary tile types, namely
CNN tiles and LSTM tiles. Both tile types share the same ReRAM-based
VMM substrate, but adopt different datapath organizations according to
the computation pattern of each branch.

\textbf{CNN tile.}
As shown in Fig.~\ref{fig:architecture}(b), the CNN tile is optimized
for convolutional and FC operations in the spatial feature extraction
path. Input feature maps and intermediate activations are stored in
local eDRAM buffers. For convolution, the input vector generation unit
converts convolution windows into vectors using the image-to-column
(im2col) method. These vectors are then applied to the DAC array and
fed into the ReRAM VMM unit, where CNN kernels or FC weights are stored
as cell conductances. The resulting analog outputs are digitized by the
ADC array, and partial sums from multiple arrays or bit slices are
combined by shift-and-add units. Finally, activation SFUs generate the
output activations or feature maps. This design follows the common
ReRAM-based convolution mapping strategy while adapting it to the
spectrogram-based CNN branch used in the depression detection
pipeline~\cite{isaac,pipelayer}.

\textbf{LSTM tile.}
As shown in Fig.~\ref{fig:architecture}(c), the LSTM tile reuses the
same ReRAM VMM primitive while incorporating additional recurrent-state
storage and gating logic for temporal modeling. The local buffers store
the current input vector $x_t$, the previous hidden state $h_{t-1}$,
and the previous cell state $C_{t-1}$. The concatenated input vector
$v_t=[x_t;h_{t-1}]$ is applied through the DAC array, while the
concatenated gate weight matrix $M=[U\;W]$ is stored in the ReRAM
crossbar arrays. In this way, the input-state and recurrent-state
matrix multiplications of each gate can be merged into a single VMM
operation~\cite{eralstm}. The digitized outputs are then processed by
sigmoid/tanh SFUs and element-wise units to compute gate activations,
cell-state updates, and hidden-state updates. The updated states $C_t$
and $h_t$ are written back to the local buffers for reuse in the next
time step.

Compared with a homogeneous tile organization, the proposed
heterogeneous tile architecture allows CNN and LSTM computations to
share the same ReRAM-based VMM substrate while preserving
module-specific datapath optimizations. CNN tiles emphasize im2col
vector generation, partial-sum accumulation, and activation
post-processing, whereas LSTM tiles emphasize local recurrent-state
buffering, element-wise state updates, and write-back across time
steps.

\begin{figure}[t]
    \centering
    \includegraphics[width=\linewidth]{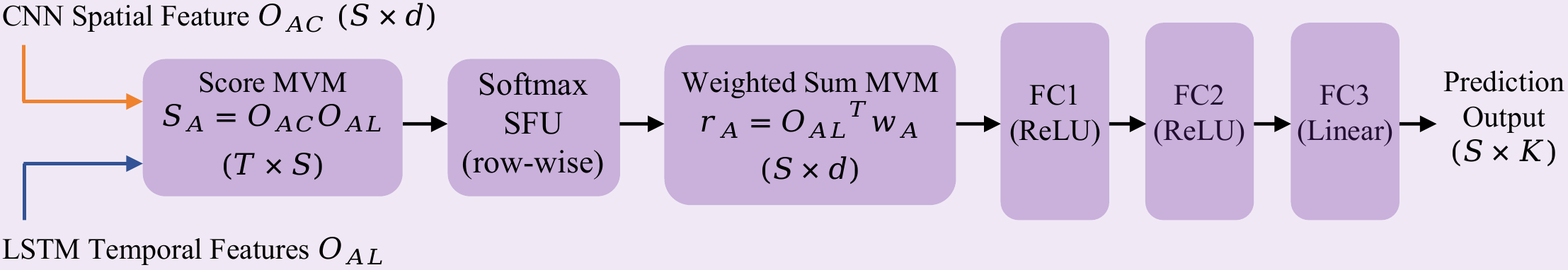}
    \caption{Attention/FC processing for ASLF representation extraction and final prediction.}
    \label{fig:attention_fc}
\end{figure}

\subsection{Attention/FC Mapping and Data Transfer}

After CNN and LSTM feature extraction, the CNN spatial feature
$O_{AC}$ and the LSTM temporal feature matrix $O_{AL}$ are transferred
through the NoC to the attention/FC processing stage, as illustrated in
Fig.~\ref{fig:architecture}(a) and Fig.~\ref{fig:attention_fc}. The
attention scores $s_A=O_{AL}O_{AC}$ and weighted aggregation
$r_A=O_{AL}^{T}w_A$ are computed by on-chip digital vector units,
while softmax is handled by an SFU. This is the attention dataflow
used in the module-level evaluation. It avoids per-inference ReRAM
programming of the input-dependent matrix $O_{AL}$ and keeps attention
processing on chip.

The fused representation $r_A$ is further processed by the FC layers for
ASLF representation extraction and final prediction. The FC weights are
mapped onto ReRAM crossbars, and the ReLU activations are handled by
SFUs, while the final linear layer produces the prediction output. The
final prediction output is then written back to local buffers as the
BDI-II score prediction for the adapted workload defined in Section II-A.
By integrating CNN
computation, LSTM computation, attention fusion, and FC processing
within a unified NoC-connected architecture, the proposed design
supports the hybrid CNN--LSTM dataflow more efficiently than treating
the spatial and temporal branches as separate accelerator subsystems.

\section{Evaluation}
\subsection{Experimental Setup and Baseline Comparison}

\textbf{Experimental setup:}
The proposed ReRAM-based CNN--LSTM accelerator is evaluated using
architectural analytical modeling and design-space exploration (DSE).
The logical architecture consists
of 64 CNN tiles and 64 LSTM tiles interconnected via a hierarchical NoC,
where each tile integrates $128 \times 128$ crossbar arrays, ADC/DAC
interfaces, eDRAM buffers, and peripheral processing units, following
the dataflow described in Sections II and III~\cite{isaac,pipelayer,eralstm}.

We adopt the 129-bin, 64-frame spectrogram input used by the audio STA
network of Niu et al.~\cite{niu2020multimodal}. In that work, the
64-frame segment length was selected through experiments on the AVEC2013
and AVEC2014 development sets; it is adopted here as the target workload
configuration.

\textbf{Module-level workload breakdown:}
The logical model uses a $129\times64$ spectrogram, 64 recurrent steps,
64-dimensional attention, a final
$64\!\rightarrow\!64\!\rightarrow\!64\!\rightarrow\!1$ MLP, and
$128\times128$ arrays. The architecture allocates 64 one-crossbar CNN tiles
and 64 one-crossbar LSTM tiles. The model excludes bit slicing,
multi-bit-cell expansion, programming, and physical ADC/DAC timing. CNN
weights are replicated across the 64 CNN tiles for concurrent patch
processing; programming and reconfiguration overheads are excluded.
Attention comprises 8,192 MAC-equivalent operations: two digital
vector/MVM-equivalent operations and one softmax/SFU operation. Because its
operands are input-dependent activations, attention is handled by digital/SFU
post-processing without per-inference ReRAM programming and contributes three
normalized cycles.

\begin{table}[!h]
\caption{Module-level logical-MVM and ReRAM-activity breakdown.}
\label{tab:module_activity_final}
\centering
\scriptsize
\setlength{\tabcolsep}{1.15pt}
\renewcommand{\arraystretch}{1.08}
\begin{tabular}{@{}lrrrrr@{}}
\toprule
\textbf{Module} & \textbf{MACs/eq.} & \textbf{Xbar ops} &
\shortstack{\textbf{Activity}\\\textbf{share}} & \textbf{Cycles} &
\textbf{$U_{\mathrm{cell}}$} \\
\midrule
Conv1          & 332,640   & 756   & 64.4\% & 12 & 2.69\% \\
Conv2          & 137,280   & 156   & 13.3\% & 3  & 5.37\% \\
CNN-FC         & 19,968    & 3     & 0.26\% & 1  & 40.63\% \\
LSTM gates     & 3,162,112 & 256   & 21.8\% & 64 & 75.39\% \\
Attention      & 8,192     & N/A   & N/A    & 3  & N/A \\
Final MLP      & 8,256     & 3     & 0.26\% & 3  & 0.39--25.00\% \\
\midrule
Total/critical & 3,668,448 & 1,174 & 100\%  & 70 & N/A \\
\bottomrule
\end{tabular}
\vspace{0.4ex}

\parbox{\columnwidth}{\scriptsize\textit{Note:} Total latency is a critical
path, not the sum of the row cycles. $U_{\mathrm{cell}}$ is mapped
weight-cell fill, not chip-area utilization. Xbar activity share indicates
ReRAM invocations, not physical or total-chip energy; attention digital/SFU
energy is excluded.}
\end{table}

The CNN branch requires 16 normalized cycles, whereas the 64-step recurrent
path requires 64. Including attention and the final MLP gives
\begin{equation}
T_{\mathrm{e2e}}=\max(16,64)+3+3=70.
\label{eq:e2e_final}
\end{equation}
LSTM gates account for 86.2\% of target MACs and form the sequential
critical-path bottleneck; additional LSTM replicas cannot remove the
64-step dependency of one sequence. Conv1 instead causes 756 of the 1,174
ReRAM invocations (64.4\%), producing the largest crossbar/peripheral access
pressure. Conv1 and Conv2 have low mapped-cell fill, while the fused
$193\times256$ LSTM matrix reaches 75.39\%.

\begin{figure}[t]
  \centering
  \includegraphics[width=\columnwidth]{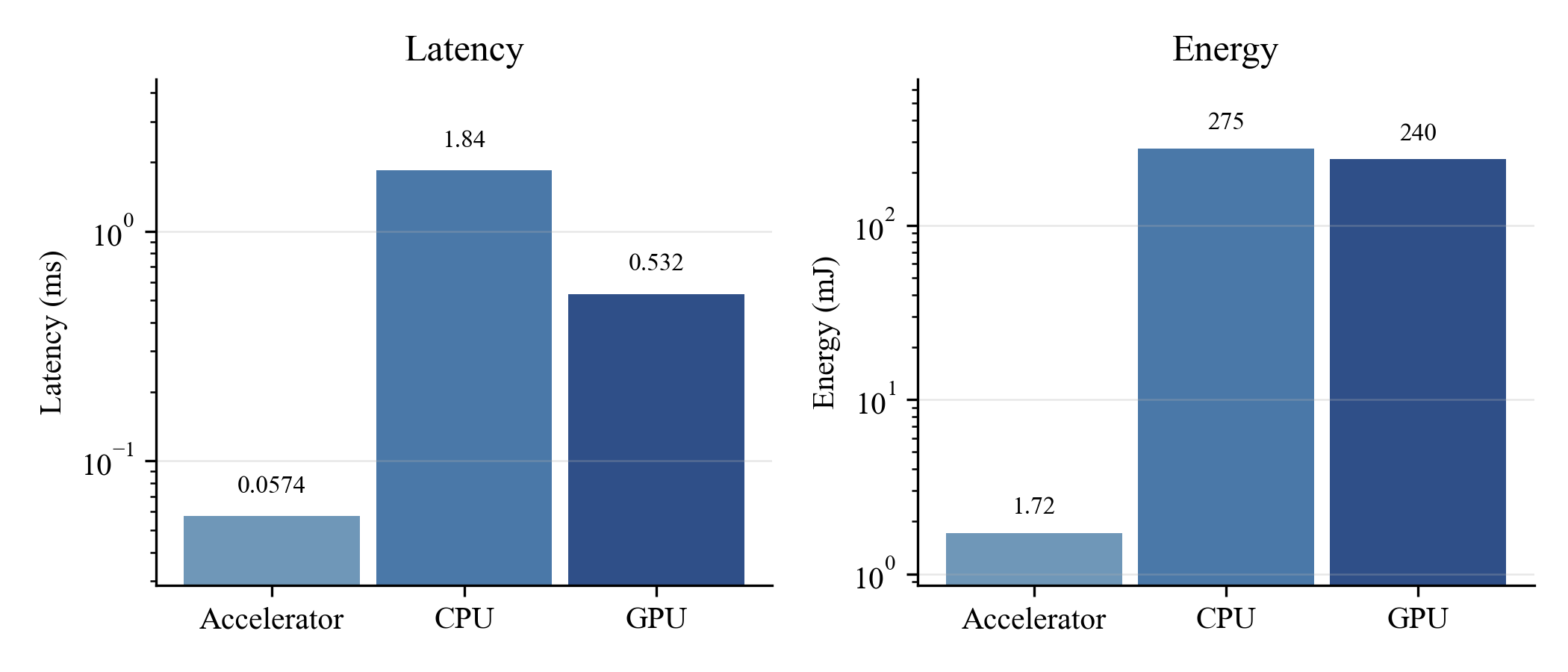}
  \caption{Latency and energy comparison of the proposed ReRAM-based
  accelerator with CPU and GPU baselines.}
  \label{fig:latency_energy}
\end{figure}

\textbf{CPU/GPU baselines:}
Fig.~\ref{fig:latency_energy} compares the latency and energy consumption
of the proposed design with CPU and GPU baselines. The proposed accelerator
achieves 0.0574 ms latency, compared with 1.84 ms on CPU and 0.532 ms on
GPU, corresponding to approximately $32.1\times$ and $9.3\times$ speedup,
respectively. In terms of energy consumption, the proposed design consumes
1.72 mJ, while the CPU and GPU baselines consume 275 mJ and 240 mJ,
respectively. This corresponds to about $160\times$ and $140\times$
energy reduction over CPU and GPU. These results indicate that the proposed
architecture effectively leverages in-memory computing to achieve both low
latency and high energy efficiency, making it well suited for real-time
inference workloads.

\textbf{Matched-workload ReRAM/PIM baselines:}
Directly reusing absolute results from prior accelerators would mix workloads,
precision, resource organization, timing scope, and peripheral assumptions.
Table~\ref{tab:matched_context_final} therefore provides mapping context
under explicitly different model scopes. The proposed-design rows use the
logical mapping and normalized cycles of Table~\ref{tab:module_activity_final};
the other rows report scoped analytical results or native capabilities.

\begin{table}[!h]
\caption{Matched-workload resource and timing context for representative
ReRAM accelerators.}
\label{tab:matched_context_final}
\centering
\scriptsize
\setlength{\tabcolsep}{1.0pt}
\renewcommand{\arraystretch}{1.09}
\begin{tabular}{@{}
>{\raggedright\arraybackslash}p{0.145\columnwidth}
>{\raggedright\arraybackslash}p{0.275\columnwidth}
>{\raggedright\arraybackslash}p{0.265\columnwidth}
>{\raggedright\arraybackslash}p{0.235\columnwidth}@{}}
\toprule
\shortstack[l]{\textbf{Branch/}\\\textbf{design}} & \textbf{Resident mapping} &
\textbf{Timing or events} & \textbf{Scope/limitation} \\
\midrule
Proposed CNN &
64 single-crossbar CNN tiles; logical array blocks per weight replica (Conv1/Conv2/FC): 1/4/3. &
16 normalized cycles: 12 + 3 + 1. &
Logical mapping; bit slicing and physical converter timing excluded. \\

ISAAC first~\cite{isaac} &
IMA/replica: 1/1/2; replicas: 8/2/1 (12 IMAs). &
185.6~$\mu$s first sample; II = 152.0~$\mu$s. &
Native 16-bit bit-serial, compute-read-only model. \\

ISAAC min-II~\cite{isaac} &
IMA/replica: 1/1/2; replicas: 9/1/1 (12 IMAs). &
198.4~$\mu$s first sample; II = 134.4~$\mu$s. &
Same compute-read-only scope. \\

Proposed LSTM &
64 single-crossbar LSTM tiles; four logical $128\times128$ array blocks per $193\times256$ weight copy. &
64 normalized cycles for 64 recurrent steps. &
Logical mapping; bit slicing and physical converter timing excluded. \\

ERA-LSTM~\cite{eralstm} &
$193\times256$ matrix; 2 tiles; 128 VMM + 128 EW arrays; tile-2 row use 50.78\%. &
Per step: 128 reads, 384 ADC, 513 DAC. $T_{\mathrm{seq}}=64(T_{\mathrm{VMM}}+T_{\mathrm{agg}}+T_{\mathrm{NL/EW}}+T_{\mathrm{comm}})$. &
Event-count estimate under our interpreted two-tile mapping; nonlinear/EW latency not quantified. \\

PUMA~\cite{puma2019} &
Programmable MVM, vector, and nonlinear ISA; target graph not benchmarked. &
Target latency requires compiler partitioning and scheduling. &
Original at 1~GHz: 577~GOPS/s/mm$^2$, 837~GOPS/s/W; not a target result. \\
\bottomrule
\end{tabular}
\vspace{0.4ex}

\parbox{\columnwidth}{\scriptsize\textit{Note:} Proposed-design cycles use
the logical model of Table~\ref{tab:module_activity_final} and are not physical
timing estimates. ISAAC timings are analytical compute-read results and exclude target-specific eDRAM, NoC, interstage
buffering, nonlinear processing, and pipeline fill/drain. ERA event counts
are derived from the interpreted dataflow. No speedup or cross-architecture
energy/area ratio is inferred from these heterogeneous scopes.}
\end{table}

The module-level analysis shows that the target workload combines high CNN
invocation pressure with a recurrent latency bottleneck. ISAAC provides a
CNN-oriented layer-resident substrate, ERA-LSTM integrates recurrent,
nonlinear, and element-wise processing, and PUMA offers broader
programmability but requires compiler-level target mapping. The proposed
architecture integrates the major CNN, LSTM, attention, and MLP stages.
Because the evaluations use different precision, resource organization,
timing scope, and peripheral models, the results provide mapping and
performance context rather than an absolute speed, energy-efficiency, or area
ranking. A cross-architecture area ranking is omitted because the proposed
digital/SFU and peripheral area has not been modeled under a common
technology.

\subsection{Hardware-Level Analysis and Robustness}
\begin{figure}[t]
    \centering
    \includegraphics[width=\columnwidth]{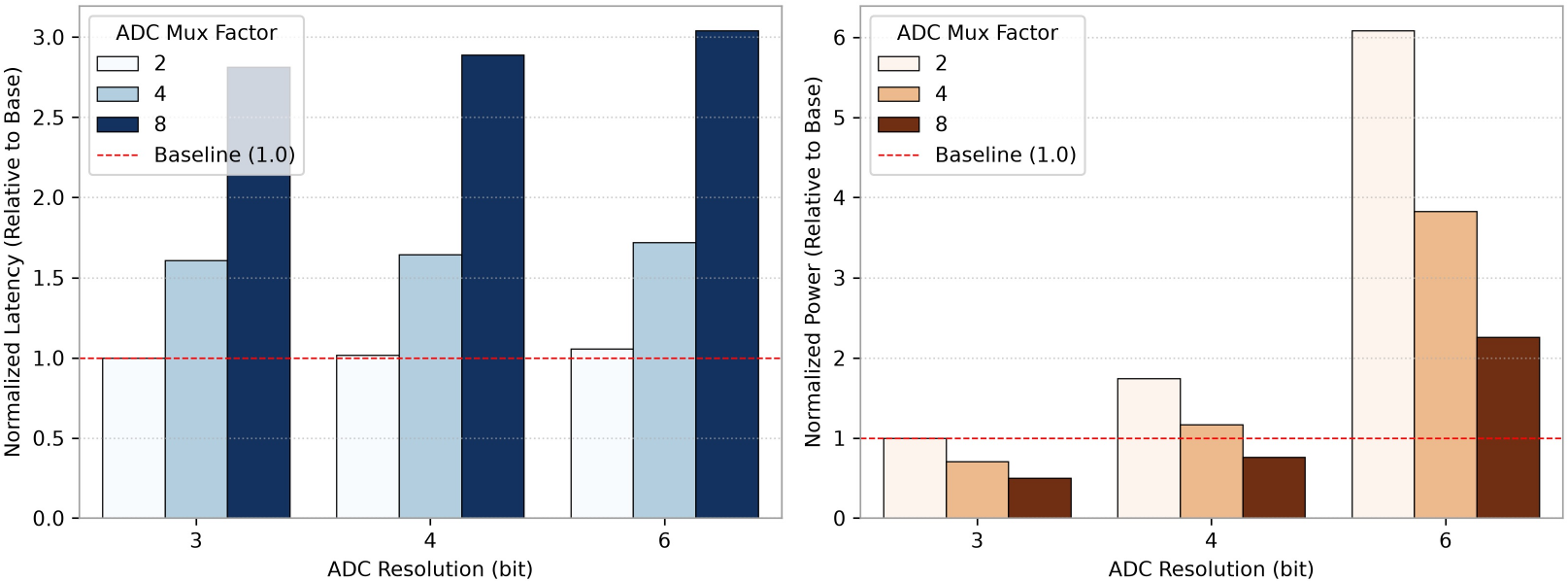}
    \caption{Sensitivity of ADC resolution and ADC mux factor:
    (a) normalized system latency and (b) normalized runtime power.}
    \label{fig:sensitivity_adc}
\end{figure}

\begin{figure}[t]
    \centering
    \includegraphics[width=0.85\columnwidth]{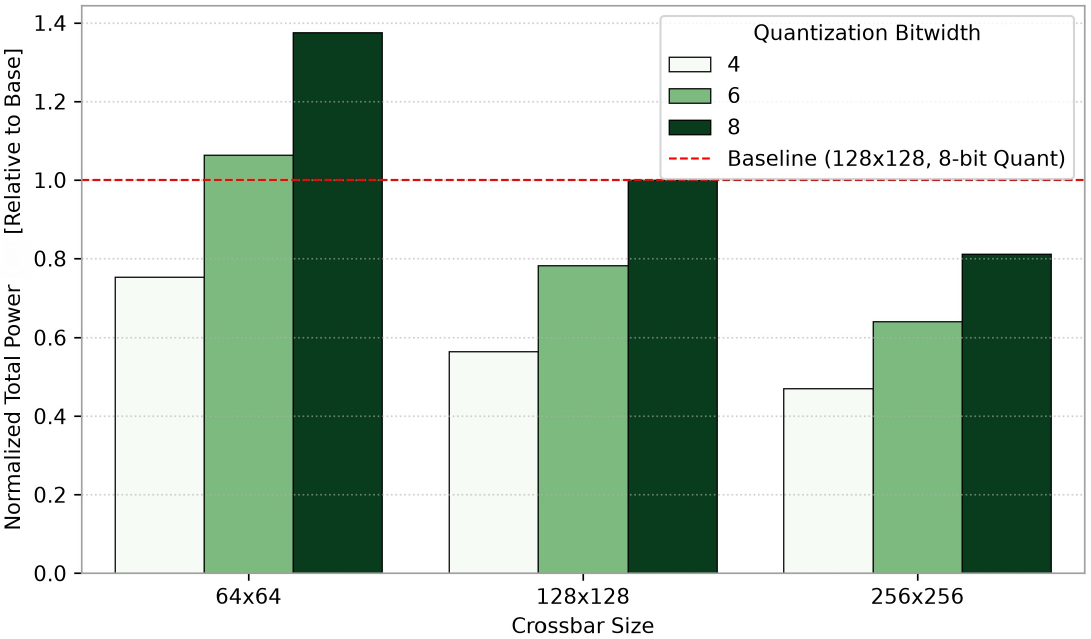}
    \caption{Sensitivity of crossbar size and quantization bitwidth on normalized power.}
    \label{fig:sensitivity_crossbar}
\end{figure}

\textbf{Parameter sensitivity:}
We conduct a design space exploration (DSE) to evaluate the impact of
architectural parameters on system latency and average runtime power.
The swept parameters include ADC resolution, ADC mux factor, crossbar
size, and quantization bitwidth. Parameters not involved in the current
sweep are fixed to the nominal settings: $128\times128$ crossbar size,
8-bit quantization, 4-bit ADC resolution, and ADC mux factor of 2. The
normalized results are reported relative to the corresponding reference
configuration indicated by the red dashed baseline.
Figure~\ref{fig:sensitivity_adc} uses a 3-bit ADC, mux-2 reference,
whereas Fig.~\ref{fig:sensitivity_crossbar} uses a 4-bit ADC, mux-2
reference. Both use $128\times128$ crossbars and 8-bit quantization.
The DSE timings and power are model estimates reported separately from
the normalized logical cycles in Table~\ref{tab:module_activity_final}.

The ADC-related sweep in Fig.~\ref{fig:sensitivity_adc} couples ADC
resolution and mux factor because both directly affect peripheral signal
conversion. The reference performance is 647.38 ns latency and 0.0452 W
average power. With 3-bit ADC resolution, increasing the mux factor from
2 to 4 and 8 increases latency by 60.5\% and 181.4\%, but reduces
average power by 29.3\% and 50.2\%, respectively. With the mux factor
fixed at 2, increasing ADC resolution from 3-bit to 4-bit and 6-bit only
increases latency by 1.9\% and 5.7\%, while average power rises by
74.5\% and 508.7\%. These results indicate that ADC muxing mainly
trades latency for power reduction, whereas ADC resolution dominates
runtime power.

For crossbar size and quantization bitwidth, the latency impact is
limited in our evaluation, so Fig.~\ref{fig:sensitivity_crossbar}
focuses on normalized average power. The reference performance is
659.69 ns latency and 0.0789 W average power. With 8-bit quantization,
shrinking the crossbar from $128\times128$ to $64\times64$ increases
average power by 37.6\%, while expanding it to $256\times256$ reduces
average power by 18.8\%. With the $128\times128$ crossbar fixed,
reducing quantization from 8-bit to 6-bit and 4-bit lowers average power
by 21.8\% and 43.5\%, respectively. These results characterize the
modeled latency and power effects of the architectural parameters.
They do not establish numerical fidelity at the corresponding ADC
resolutions or quantization bitwidths. Task-level MAE/RMSE and the
accuracy effects of ADC conversion and weight quantization remain
to be evaluated.

\begin{figure}[t]
    \centering
    \begin{minipage}{0.49\columnwidth}
        \centering
        \includegraphics[width=\linewidth]{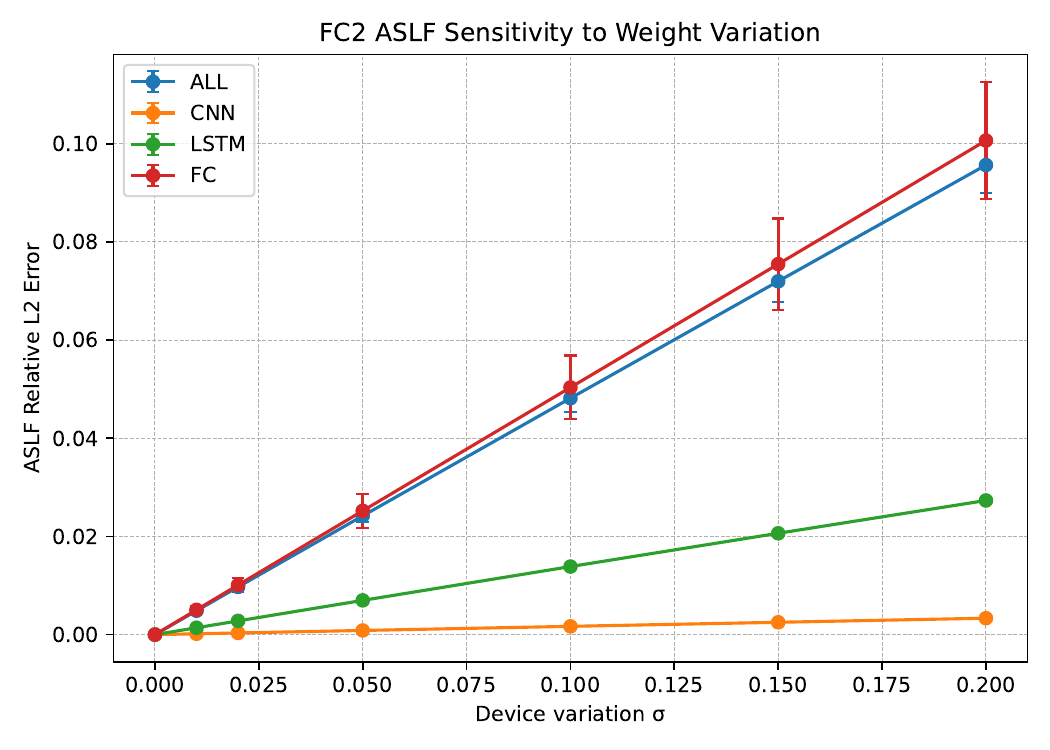}
        \centerline{(a) Weight variation}
    \end{minipage}
    \hfill
    \begin{minipage}{0.49\columnwidth}
        \centering
        \includegraphics[width=\linewidth]{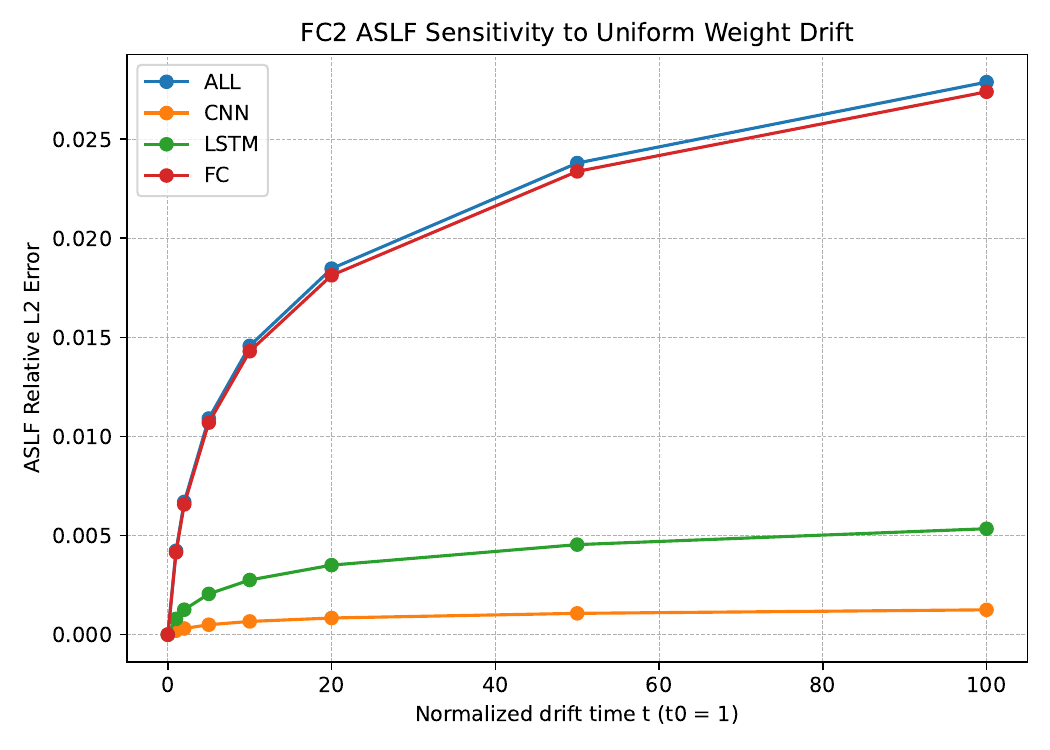}
        \centerline{(b) Uniform weight drift}
    \end{minipage}
    \caption{ASLF sensitivity of an untrained CNN--LSTM pipeline
    to simplified weight perturbations on synthetic inputs.}
    \label{fig:robustness}
\end{figure}

\textbf{Representation-Level Sensitivity Analysis:}
Motivated by the non-idealities of analogue-memory computing discussed in
prior work~\cite{rram_survey,ambrogio2018nature}, we evaluate ASLF
sensitivity using simplified weight-variation and drift models.
The tested network follows the CNN--LSTM attention structure defined in
Section~II-A, with a ReLU after the CNN projection and ASLF extracted
after FC2/ReLU, before the scalar linear FC3 head.
The model is randomly initialized with seed 42 and is not trained.
A fixed batch of 128 synthetic standard-normal inputs of shape
$1\times129\times64$ is reused for all settings.
The tests characterize representation sensitivity, not task-level
prediction accuracy or the robustness of a trained depression detector.

The CNN group includes the convolutional and projection weights, the
LSTM group includes the input and recurrent weights, and the FC group
includes FC1 and FC2 weights. ALL denotes the union of these tested
groups. Biases and the downstream FC3 head are not perturbed.
Weight variation is modeled as:
\begin{equation}
    W' = W\cdot\left(1+\mathcal{N}(0,\sigma^2)\right).
\end{equation}
Independent Gaussian samples are drawn for the selected weight elements.
We sweep $\sigma\in\{0,0.01,0.02,0.05,0.10,0.15,0.20\}$ using
five noise seeds (0--4) per setting, with the clean weights and inputs fixed.
Uniform weight drift is modeled as:
\begin{equation}
    W(t) = W_0\left(1-\alpha\log\left(1+\frac{t}{t_0}\right)\right).
\end{equation}
We use $\alpha=0.01$, $t_0=1$, the natural logarithm, and
$t\in\{0,1,2,5,10,20,50,100\}$. Here, $t$ is a normalized time
index, not a physical retention time. Each drift setting is evaluated
once because the scaling is deterministic. At $t=100$, the selected
weights are multiplied by approximately 0.95385. This common scaling
does not model device-to-device drift variability or calibrated retention.

Let $a$ and $a'$ denote the vectorized clean and perturbed ASLF outputs
for the entire $128\times64$ batch. Their relative deviation is:
\begin{equation}
    E_{\mathrm{rel}} =
    \frac{\|a'-a\|_2}{\|a\|_2+\epsilon},
    \qquad \epsilon=10^{-12}.
\end{equation}
The norm is computed before averaging across noise trials; signed
residuals are not averaged before taking the norm. Figure~\ref{fig:robustness}(a)
shows the mean over five noise trials, with error bars indicating their
sample standard deviation. Figure~\ref{fig:robustness}(b) reports the
deterministic drift results without trial error bars.

At $\sigma=0.20$, the mean relative ASLF errors for CNN, LSTM, FC,
and ALL perturbations are $3.34\times10^{-3}$, $2.73\times10^{-2}$,
$1.01\times10^{-1}$, and $9.56\times10^{-2}$, respectively.
At $t=100$, the corresponding drift errors are $1.25\times10^{-3}$,
$5.34\times10^{-3}$, $2.74\times10^{-2}$, and $2.79\times10^{-2}$.
Among the individually perturbed groups, both experiments show the
ordering FC $>$ LSTM $>$ CNN for the measured deviation. This ordering
is specific to the tested initialization, synthetic inputs, and FC2
observation point; it is not a general ranking of module robustness or
final prediction accuracy. The tests do not evaluate ADC conversion,
weight quantization, or trained-model MAE/RMSE.

\section{Conclusion}

We presented a unified ReRAM-based accelerator for CNN--LSTM inference
in speech-based depression detection. The proposed architecture integrates convolutional, recurrent, attention, and MLP stages within a
crossbar-based PIM framework, reducing intermediate data movement and
supporting the hybrid spatial--temporal dataflow. Evaluation results show
$32.1\times$ and $9.3\times$ latency reduction over CPU and GPU baselines, with about $160\times$ and $140\times$ energy reduction, respectively. Module-level and matched-workload analyses further provide context against representative ReRAM/PIM accelerators. Sensitivity analysis identifies ADC resolution, mux factor, crossbar size, and quantization bitwidth as important design trade-offs. Simplified weight-perturbation tests characterize ASLF sensitivity in an untrained pipeline with synthetic inputs, without establishing task-level robustness. Future work will evaluate trained models on real depression datasets, including the task-level effects of quantization, ADC conversion, and device non-idealities.

\section*{Acknowledgment}

This work is supported by the National Natural Science Foundation of China
(62474144, 62501489 and 62174053), the Centrally Guided Local Science and
Technology Development Fund (2024ZYZX4025), the Basic Research Program of
Jiangsu (BK20250465), the Frontier Technologies R\&D Program of Jiangsu
(BF2025031), the Suzhou Industrial Park MEMS Advanced High-Performance
Sensor Chip Technology Transfer Platform (CXZ2024201), the Jiangsu Data
Science and Cognitive Computational Engineering Research Centre, the
XJTLU AI University Research Centre, the XJTLU Advanced Materials Research
Center and the XJTLU Research Development Fund (RDF-24-02-017).

\vspace{6pt}
\bibliographystyle{IEEEtran}
\bibliography{refs}


\end{document}